\documentclass[%
 reprint,
superscriptaddress,
showpacs,
 amsmath,amssymb,
 aps,
 pra,
]{revtex4-1}

\usepackage{relsize}
\usepackage{graphicx}
\usepackage{dcolumn}
\usepackage{bm}
\usepackage{color}
\usepackage{hyperref}

\newcommand{\beq}{\begin{eqnarray}}
\newcommand{\eeq}{\end{eqnarray}}
\newcommand{\bea}{\begin{align}}
\newcommand{\eea}{\end{align}}
\newcommand{\im}{\mathrm{i}}
\newcommand{\del}{\partial}
\newcommand{\de}{\mathrm{d}}

\newcommand{\n}{\mathrm{n}}

\begin{document}

\preprint{APS/123-QED}

\title{Fully general time-dependent multiconfiguration self-consistent-field method for the electron-nuclear Dynamics}

\author{Ryoji Anzaki}
\email {anzaki@atto.t.u-tokyo.ac.jp}

\affiliation{Department of Nuclear Engineering and Management, Graduate School of Engineering, The University of Tokyo, 7-3-1 Hongo, Bunkyo-ku, Tokyo 113-8656, Japan}

\author{Takeshi Sato}
\affiliation{Department of Nuclear Engineering and Management, Graduate School of Engineering, The University of Tokyo, 7-3-1 Hongo, Bunkyo-ku, Tokyo 113-8656, Japan}
\affiliation{Photon Science Center, Graduate School of Engineering, The University of Tokyo, 7-3-1 Hongo, Bunkyo-ku, Tokyo 113-8656, Japan}

\author{Kenichi L. Ishikawa}
\affiliation{Department of Nuclear Engineering and Management, Graduate School of Engineering, The University of Tokyo, 7-3-1 Hongo, Bunkyo-ku, Tokyo 113-8656, Japan}
\affiliation{Photon Science Center, Graduate School of Engineering, The University of Tokyo, 7-3-1 Hongo, Bunkyo-ku, Tokyo 113-8656, Japan}

\date{\today}

\begin{abstract}
We present the fully general time-dependent  multiconfiguration self-consistent-field method to describe the dynamics of a system consisting of arbitrary different kinds and numbers of interacting fermions and bosons. 
The total wave function is expressed as a superposition of different configurations constructed from time-dependent spin-orbitals prepared for each particle kind.
We derive equations of motion followed by configuration-interaction (CI) coefficients and spin-orbitals for general, not restricted to full-CI, configuration spaces.
The present method provides a flexible framework for the first-principles theoretical study of, e.g., correlated multielectron and multinucleus quantum dynamics in general molecules induced by intense laser fields and attosecond light pulses.

\end{abstract}

\maketitle

\section{Introduction}


We are now witnessing rapid progress in ultrashort intense light sources in different spectral ranges such as terahertz radiation, optical-parametric-chirped-pulse-amplification mid-infrared lasers, high-harmonic extreme-ultraviolet (XUV) pulses, and XUV/x-ray free-electron lasers. 
These technological advances have triggered various research activities, including attosecond science \cite{Agostini2004RPP, Krausz2009RMP, Gallmann2012ARPC}, with a goal to directly measure and, ultimately, control electron and nuclear motion in atoms and molecules.

{\it Ab initio} simulations of the electronic and nuclear dynamics in atoms and molecules remain a challenge. The multiconfiguration time-dependent Hartree-Fock (MCTDHF) method \cite{Kato2004CPL, Caillat1} has been developed for the investigation of multielectron dynamics in strong and/or ultrashort laser fields \cite{Ishikawa2015JSTQE}. 
In this approach, the time-dependent total electronic wave function $\Psi (t)$ is expressed as a superposition of different Slater determinants $\Phi_{\pmb{I}}(t)$,
\begin{equation}
\label{eq:MCTDHF-electron}
	\Psi (t) = \sum_{\pmb{I}} \Phi_{\pmb{I}}(t) C_{\pmb{I}}(t),
\end{equation}
where $C_{\pmb{I}}(t)$ is the configuration-interaction (CI) coefficients. Both $\{C_{\pmb{I}}(t)\}$ and the spin-orbitals constituting $\{\Phi_{\pmb{I}}(t)\}$ are allowed to vary in time.
In the community of high-field phenomena and attosecond physics, the term MCTDHF is conventionally used for the full-CI case, in which the sum in Eq.~(\ref{eq:MCTDHF-electron}) runs over all the possible ways to distribute the electrons among a given number of spin-orbitals.
On the other hand, also under active development are variants without the restriction to the full-CI expansion, generically referred to as the time-dependent multiconfiguration self-consistent-field (TD-MCSCF) methods hereafter. The representative examples include the time-dependent complete-active-space self-consistent-field \cite{Sato1,Sato2016PRA}, the time-dependent restricted-active-space self-consistent-field \cite{Miyagi2014PRA}, and the time-dependent occupation-restricted multiple active-space (TD-ORMAS) \cite{Sato2} methods. These allow a compact and computationally less demanding description of the multielectron dynamics, without sacrificing accuracy. In particular, the TD-ORMAS can treat arbitrary CI expansions of the form Eq.~(\ref{eq:MCTDHF-electron}) in principle.

Among successful approaches for nuclear dynamics is the multiconfiguration time-dependent Hartree (MCTDH) method \cite{Meyer1990CPL}. Developed for systems consisting of distinguishable particles, this method expresses the time-dependent total nuclear wave function as a superposition similar to Eq.~(\ref{eq:MCTDHF-electron}) but that of Hartree products. The other way around, the MCTDHF can be viewed as an extension of the MCTDH to fermions. By hybridizing the MCTDHF for electrons and the MCTDH for nuclei, one can construct a multiconfiguration electron-nuclear dynamics (MCEND) method \cite{Ulusoy1} to describe the non-Born-Oppenheimer coupled dynamics. Nuclei forming molecules are, however, indistinguishable particles, either fermions or bosons. 

%

In this Paper, further stepping forward in this direction, we present a
fully general TD-MCSCF method for a system comprising of arbitrary
different kinds and numbers of interacting fermions and bosons. Treating
all the constituent particles on an equal footing, we expand the total
wave function in terms of configurations of the whole system [see
Eq.~(\ref{eq:MCTDHF ansatz}) below], rather than considering
configurations of each particle kind separately as in
Ref.~\cite{Alon1}. 
Thus, based on the time-dependent variational principle, we derive the
equations of motion (EOM) of CI coefficients and spin-orbitals for
general configuration spaces, not restricted to full-CI. 

This paper is organized as follows. 
Section \ref{sec:definitions} introduces our
TD-MCSCF ansatz for many-particle systems composed of different kinds of
fermions and bosons, and also defines the target Hamiltonian considered
in this work. In Sec.~\ref{sec:EOM}, we
derive the general equations of motion, based on the time-dependent
variational principle. Explicit working equations for a molecule interacting with an external laser field are
shown in Sec.~\ref{sec:molecule}. Concluding remarks are given in
Sec.~\ref{sec:summary}. 


\section{Definition of the problem}\label{sec:definitions}
\subsection{TD-MCSCF ansatz}
We consider a quantum mechanical many-body system with
$K$ kinds of fermions or bosons.
The subsystem of kind $\alpha$ consists of $N_{\alpha}$ identical particles. 
Thus, there are $N = \sum_{\alpha =1}^{K} N_\alpha$ particles in
whole. For notational brevity, we call such a system an
$\pmb{N}$-particle system, where the array of integers 
$\pmb{N}=(N_1N_2\cdots N_K)$ carries information of both
particle kinds and number of particles in each kind.

Let us define, for each kind of particles, the complete orthonormal set of spin-orbitals
$\{\chi^{(\alpha)}_{\mu_\alpha}(t): \mu_\alpha \in \Omega_{\alpha}\}$, which spans the one-particle Hilbert space $\Omega_\alpha$, 
and are time-dependent in general.
Then the $\pmb{N}$-particle Hilbert space is spanned by
\beq\label{eq:configuration1}
\Phi_{\pmb{I}}(t) = \Phi^{(1)}_{\pmb{I}_1}(t)\otimes \Phi^{(2)}_{\pmb{I}_2}(t)\otimes\cdots\otimes \Phi^{(K)}_{\pmb{I}_K}(t),
\eeq
where
$\Phi^{(\alpha)}_{\pmb{I}_\alpha}(t)$ is a determinant (or parmanent) of $\alpha$-kind
fermions (or bosons), consisting of $N_\alpha$ spin-orbitals chosen from
$\{\chi^{(\alpha)}_{\mu_\alpha}\}$.
We call $\Phi_{\pmb{I}}(t)$ the $\pmb{I}$'s configuration, where $\pmb{I}=\pmb{I}_1\pmb{I}_2\cdots \pmb{I}_K$
is considered, at the moment, to collectively label the chosen spin-orbitals.
The objective of this paper is to formulate the TD-MCSCF theory of the
$\pmb{N}$-particle system within the ansatz of total wavefunction
analogous to that for electronic system, Eq.~(\ref{eq:MCTDHF-electron}),
but using the configurations of Eq.~(\ref{eq:configuration1}).

For rigorous and compact presentation of theory, we resort to the second
quantization formulation by introducing creation and annihilation operators
$\{\hat{c}^{(\alpha)\dagger}_{\mu_\alpha},\hat{c}^{(\alpha)}_{\mu_\alpha}\}$ associated to
$\{\chi^{(\alpha)}_{\mu_\alpha}\}$. 
These operators obey the (anti-)commutation relations of bosons (fermions),
\beq\label{eq:statisticsb}
[\hat{c}^{(\alpha)}_{\mu_\alpha},\hat{c}^{(\alpha)}_{\nu_\alpha}] =
[\hat{c}^{(\alpha)\dagger}_{\mu_\alpha},\hat{c}^{(\alpha)\dagger}_{\nu_\alpha}]
= 0, \quad
[\hat{c}^{(\alpha)}_{\mu_\alpha},\hat{c}^{(\alpha)\dagger}_{\nu_\alpha}]
= \delta^{\mu_\alpha}_{\nu_\alpha}, 
\eeq
for bosons, where $[\hat{a},\hat{b}]=\hat{a}\hat{b}-\hat{b}\hat{a}$, and
\begin{align}
\label{eq:statisticsf}
\{\hat{c}^{(\alpha)}_{\mu_\alpha},\hat{c}^{(\alpha)}_{\nu_\alpha}\} =
\{\hat{c}^{(\alpha)\dagger}_{\mu_\alpha},\hat{c}^{(\alpha)\dagger}_{\nu_\alpha}\}
= 0, \quad
\{\hat{c}^{(\alpha)}_{\mu_\alpha},\hat{c}^{(\alpha)\dagger}_{\nu_\alpha}\}
= \delta^{\mu_\alpha}_{\nu_\alpha},
\end{align}
for fermions, where $\{\hat{a},\hat{b}\}=\hat{a}\hat{b}+\hat{b}\hat{a}$.

Within the TD-MCSCF ansatz, 
the complete set of spin-orbitals $\{\chi_{\mu_\alpha}^{(\alpha)}(t)\}$ is split
into $n_\alpha$ ($\geq N_\alpha$) {\it occupied} spin-orbitals
$\{\chi_{i_\alpha}^{(\alpha)}(t): i_\alpha =1,2,\cdots ,n_\alpha\}$ 
and remaining {\it
virtual} spin-orbitals $\{\chi_{a_\alpha}^{(\alpha)}(t): a_\alpha =n+1,n+2\cdots\}$.
We call the subspace of $\Omega_\alpha$ spanned by occupied
spin-orbitals the occupied spin-orbital space $\Omega^{occ}_\alpha$, and that
spanned by virtual spin-orbitals the virtual spin-orbital space
$\Omega^{vir}_\alpha$, where $\Omega_\alpha =\Omega^{occ}_\alpha\oplus\Omega^{vir}_\alpha$.
The total state $\Psi(t)$ is expressed as a superposition of 
configurations $\Phi_{\pmb{I}}(t)$ of Eq.~(\ref{eq:configuration1}), but
constructed from occupied spin-orbitals only.
Thus we write
\beq
\label{eq:MCTDHF ansatz}
|\Psi(t)\rangle = \sum_{\pmb{I}}C_{\pmb{I}}(t)|\pmb{I}(t)\rangle,
\eeq
where $C_{\pmb{I}}(t)$ is the CI coefficient, and $|\pmb{I}(t)\rangle$ is
the occupation number representation of the configuration $\Phi_{\pmb{I}}$,
\beq
|\pmb{I}(t)\rangle = |\pmb{I}_1(t)\rangle\otimes|\pmb{I}_2(t)\rangle\otimes\dots \otimes|\pmb{I}_K(t)\rangle,
\eeq
\begin{align}
\label{eq:configuration2_each}
|\pmb{I}_\alpha\rangle = \frac{1}{\prod_{j_\alpha =1}^{n_\alpha} I_{\alpha,j_\alpha}!}
[\hat{c}^\dagger_{1}]^{I_{\alpha,1}}
[\hat{c}^\dagger_{2}]^{I_{\alpha,2}}\cdots
[\hat{c}^\dagger_{\mu}]^{I_{\alpha,n_\alpha}}|vac\rangle.
\end{align}
Now $\pmb{I}_\alpha=\pmb{I}_{\alpha,1}\pmb{I}_{\alpha,2}\cdots \pmb{I}_{\alpha,n_\alpha}$
is (rigorously) reinterpretted as an integer array, satisfying
$\sum_{i_\alpha =1}^{n_\alpha}I_{\alpha,i_\alpha} = N_\alpha$. Note 
that $I_{\alpha,i_\alpha} \in \{0,1\}$ for fermions.
Here and in what follows, we use indices
$i_\alpha,j_\alpha,k_\alpha,...$ for occupied ($\Omega^{occ}_\alpha$), $a_\alpha,b_\alpha,c_\alpha,...$
for virtual ($\Omega^{vir}_\alpha$), and {$\mu_\alpha,\nu_\alpha,\kappa_\alpha,\tau_\alpha,...$}
for general ($\Omega_\alpha$) spin-orbitals of kind $\alpha$. The indices
$p_\alpha,q_\alpha$ will be used for numbering the coordinates.

It should be noted that we do not restrict the expansion
Eq.~(\ref{eq:MCTDHF ansatz}) to the full-CI one. It should also be
noticed that occupied configurations are specified in terms of the whole
system rather than in terms of each particle kind separately as
\cite{Alon1},
\begin{align}
|\Psi(t)\rangle &= \sum_{{\pmb{I}}_1} \sum_{\pmb{I}_2}\cdots\sum_{\pmb{I}_K} \nonumber\\
&C_{\pmb{I}_1\pmb{I}_2\cdots\pmb{I}_K}(t)|\pmb{I}_1(t)\rangle\otimes|\pmb{I}_2(t)\rangle\otimes\cdots\otimes|\pmb{I}_K(t)\rangle.
\label{eq:MCTDHF ansatz we do not use}
\end{align}
with $\pmb{I}_\alpha, \pmb{I}_\beta, \pmb{I}_\gamma, \cdots$ being the configuration of particle kind $\alpha, \beta, \gamma, \cdots$, respectively, and $C_{\pmb{I}_\alpha \pmb{I}_\beta \pmb{I}_\gamma \cdots}(t)$ the CI coefficient.
Our approach allows a highly flexible choice of CI space, e.g., including up to double excitation \cite{Sato2} regardless of particle kind, thereby enabling proper account of correlation between different kinds of particles while suppressing computational cost.

\subsection{Target Hamiltonian}
In this article, we consider the Hamiltonian of an $\pmb{N}$-particle
system composed of up to $M$-body terms,
\begin{eqnarray}
H = H_1+H_2+\cdots +H_{M}, \quad M \leq N.
\end{eqnarray}
The Hamiltonian is explicitly time-dependent in general, but the time
argument $t$ is dropped in this section for simplicity. 
Here, the $m$-body Hamiltonian is assumed to be given explicitly in terms of
the coordinates (and momenta, see below) in a general sense characterizing  $m$ particles (or degrees of freedom), and symmetric
under exchange of coordinates among particles of the same kind.
One-particle Hamiltonian, e.g., is written as
\begin{eqnarray}
H_1 = \sum_{\alpha =1}^K \sum_{p_\alpha =1}^{N_\alpha}H_{\alpha}(x_{\alpha ,p_\alpha},x^\prime_{\alpha ,p_\alpha}),
\end{eqnarray}
where the non-local form allows to describe the momentum 
dependence of the Hamiltonian,
and two-body interaction is generally given by
\begin{align}
H_2 &= \sum_{\alpha =1}^K \sum_{p_\alpha =1}^{N_\alpha}\sum_{q_\alpha >p_\alpha}^{N_\alpha} H_{\alpha\alpha}(x_{\alpha ,p_\alpha},x_{\beta
,q_\beta},x^\prime_{\alpha ,p_\alpha},x^\prime_{\beta ,q_\beta}) \nonumber \\ &+
\sum_{\alpha =1}^K \sum_{\beta >\alpha}^K \sum_{p_\alpha=1}^{N_\alpha} \sum_{q_\alpha = 1}^{N_\beta}H_{\alpha\beta}(x_{\alpha ,p_\alpha},x_{\beta ,q_\beta},x^\prime_{\alpha ,p_\alpha},x^\prime_{\beta ,q_\beta}).
\end{align}
The reasons why we here consider the (non-local) higher-than-two body
terms, which will not actually be used in Sec.~\ref{sec:molecule}, are (1) that such
form is used in multiconfiguration Hartree (MCH) method for
distinguishable particles, and (2) their possible appearance
upon coordinate transformations, or in the effort of removing
translational and rotational degrees of freedom\cite{Nakai,NakaiComm}.

The Hamiltonian is equivalently expressed in the second quantization
formalism as
\begin{eqnarray}
\hat{H}_m = \sum_{m_1,\cdots ,m_K} \hat{H}_{m_1,\cdots ,m_K}
= \sum_{\pmb{m}} \hat{H}_{\pmb{m}}, 
\end{eqnarray}
where the net $m$-body Hamiltonian is further classified into those
contributions $\hat{H}_{\pmb{m}}$, hereafter called $\pmb{m}$-body Hamiltonian, involving $m_\alpha$ particles of the kind $\alpha$,
($0 \leq m_\alpha \leq N_\alpha$, $\sum_{\alpha = 1}^K m_\alpha = m$),
\begin{align}
\hat{H}_{\pmb{m}} &=
\sum_{\pmb{\mu}_1}\cdots\sum_{\pmb{\mu}_K}
\sum_{\pmb{\nu}_1}\cdots\sum_{\pmb{\nu}_K}
(H_{\pmb{m}})^{\pmb{\mu}_1\cdots\pmb{\mu}_K}_{\pmb{\nu}_1\cdots\pmb{\nu}_K}
\hat{E}^{\pmb{\mu}_1\cdots\pmb{\mu}_1}_{\pmb{\nu}_1\cdots\pmb{\nu}_K} \nonumber \\
&=\sum_{\pmb{\mu}\pmb{\nu}}(H_{\pmb{m}})^{\pmb{\mu}}_{\pmb{\nu}} \hat{E}^{\pmb{\mu}}_{\pmb{\nu}}, 
\end{align}
where $\pmb{\mu} = (\pmb{\mu}_1\pmb{\mu}_2\cdots\pmb{\mu}_K)$, and 
$\pmb{\mu}_\alpha = (\mu_{\alpha ,1}\mu_{\alpha ,2}\cdots\mu_{\alpha
,m_\alpha})$ indexes the set of spin-orbitals to represent $m_\alpha$
particles in the Hamiltonian.
$\hat{E}^{\pmb{\mu}}_{\pmb{\nu}}$ is the $\pmb{m}$-particle replacement operator
$\hat{E}^{\pmb{\mu}}_{\pmb{\nu}}
=(\hat{E}_1)^{\pmb{\mu}_1}_{\pmb{\nu}_1}\cdots(\hat{E}_K)^{\pmb{\mu}_K}_{\pmb{\nu}_K}$, with
\begin{align}
(\hat{E}_\alpha)^{\pmb{\mu}_\alpha}_{\pmb{\nu}_\alpha} =
\hat{c}^{(\alpha)\dagger}_{\mu_{\alpha,1}}
\hat{c}^{(\alpha)\dagger}_{\mu_{\alpha,2}}\cdots
\hat{c}^{(\alpha)\dagger}_{\mu_{\alpha,m_\alpha}}
\hat{c}^{(\alpha)}_{\nu_{\alpha,m_\alpha}}\cdots
\hat{c}^{(\alpha)}_{\nu_{\alpha,2}}
\hat{c}^{(\alpha)}_{\nu_{\alpha,1}},
\end{align}
and $(H_{\pmb{m}})^{\pmb{\mu}}_{\pmb{\nu}}$ is given by
\begin{align}
\label{eq:hammat}
(H_{\pmb{m}})^{\pmb{\mu}}_{\pmb{\nu}} = \frac{1}{\prod_{\alpha =1}^K m_\alpha !} \sum_{\pmb{\mu}\pmb{\nu}}\int\de\pmb{x}\de\pmb{x}^\prime
\varphi_{\pmb{\mu}}^*(\pmb{x})H_{\pmb{m}}(\pmb{x},\pmb{x}^\prime)\varphi_{\pmb{\nu}}(\pmb{x}^\prime),
\end{align}
where $\pmb{x}=(\pmb{x}_1\pmb{x}_2\cdots\pmb{x}_K)$,
$\pmb{x}_\alpha=(x_{\alpha ,1}x_{\alpha ,2}\cdots x_{\alpha ,m_\alpha})$
is the set of $m_\alpha$ coordinates of particle $\alpha$, and
\begin{align}
\varphi_{\pmb{\mu}}(\pmb{x})&=\prod_{\alpha =1}^K\varphi^{(\alpha)}_{\pmb{\mu}_\alpha}(\pmb{x}_\alpha) \nonumber\\
&=\prod_{\alpha =1}^K
\chi^{(\alpha)}_{\mu_{\alpha ,1}}(x_{\alpha ,1})\chi^{(\alpha)}_{\mu_{\alpha ,2}}(x_{\alpha ,2})\cdots\chi^{(\alpha)}_{\mu_{\alpha ,m_\alpha}}(x_{\alpha ,m_\alpha}).
\end{align}

For the later discussion, we define the $\pmb{m}$-body reduced density
matrix (RDM) as
\beq\label{eq:mRDM}
(\rho_{\pmb{m}})^{\pmb{\mu}}_{\pmb{\nu}} = \langle\Psi|\hat{E}^{\pmb{\nu}}_{\pmb{\mu}}|\Psi\rangle.
\eeq
One- and two-particle RDMs are also denoted as
\beq\label{eq:12RDM}
&&(\rho_\alpha)^{\mu_\alpha}_{\nu_\alpha}=\langle\Psi|(\hat{E}_\alpha)^{\nu_\alpha}_{\mu_\alpha}|\Psi\rangle
=(\rho_{0_1\cdots1_\alpha\cdots0_K})^{\mu_\alpha}_{\nu_\alpha},\nonumber \\
&&(\rho_{\alpha\alpha})^{\mu_\alpha\gamma_\alpha}_{\nu_\alpha\lambda_\alpha}=
\langle\Psi|(\hat{E}_\alpha)_{\mu_\alpha\gamma_\alpha}^{\nu_\alpha\lambda_\alpha}|\Psi\rangle =
(\rho_{0_1\cdots 2_\alpha\cdots0_K})^{\mu_\alpha\gamma_\alpha}_{\nu_\alpha\lambda_\alpha}, \\
&&(\rho_{\alpha\beta})^{\mu_\alpha\gamma_\beta}_{\nu_\alpha\lambda_\beta} =
\langle\Psi|(\hat{E}_\alpha)_{\mu_\alpha}^{\nu_\alpha}(\hat{E}_\alpha)_{\gamma_\beta}^{\lambda_\beta}|\Psi\rangle =
(\rho_{0_1\cdots 1_\alpha\cdot\cdot
1_\beta\cdots0_K})^{\mu_\alpha\gamma_\beta}_{\nu_\alpha\lambda_\beta},\nonumber
\eeq
with $\beta\neq\alpha$.

\section{Equations of Motion}
\label{sec:EOM}

In this section, we derive the EOMs for the CI coefficients and
spin-orbitals by imposing the time-dependent variational
principle\cite{Dalgarno1,Lowdin1,Moccia1} on our TD-MCSCF ansatz.
We require the action integral
\beq\label{Var1}
\quad S = \int_{t_0}^{t_1}\de t\langle\Psi|(\hat{H}-\im\del_t)|\Psi\rangle,
\eeq 
to be stationary, $\delta S = 0$,
with respect to the variation of the total wavefunction $\delta\Psi$
within our TD-MCSCF ansatz Eq.~(\ref{eq:MCTDHF ansatz}), subject
to the boundary conditions $\delta\Psi(t_0)=\delta\Psi(t_1)=0$. 
To this end, let us introduce anti-Hermitian matrices $\Delta_\alpha$ and $X_\alpha$ as,
\begin{align}\label{defgf}
\langle\chi^{(\alpha)}_{\mu_\alpha}|\delta\chi^{(\alpha)}_{\nu_\alpha}\rangle = (\Delta_\alpha)^{\mu_\alpha}_{\nu_\alpha}, \quad \langle\chi^{(\alpha)}_{\mu_\alpha}|\dot{\chi}^{(\alpha)}_{\nu_\alpha}\rangle = (X_\alpha)^{\mu_\alpha}_{\nu_\alpha}.
\end{align}
(Recall that indices $\mu_\alpha ,\nu_\alpha$ refer to both occupied and
virtual spin-orbitals.)
%
We also define,
\begin{align}
\hat{\Delta} =
\sum_\alpha\sum_{\mu_\alpha\nu_\alpha}(\Delta_\alpha)^{\mu_\alpha}_{\nu_\alpha}(\hat{E}_\alpha)^{\mu_\alpha}_{\nu_\alpha},
\quad 
\hat{X} = \sum_\alpha\sum_{\mu_\alpha\nu_\alpha}(X_\alpha)^{\mu_\alpha}_{\nu_\alpha}(\hat{E}_\alpha)^{\mu_\alpha}_{\nu_\alpha},
\end{align}
with which orthonormality-conserving spin-orbital variations and time
derivatives can be written as
\beq
|\delta\chi_{\mu_\alpha}^{(\alpha)}\rangle=\hat{\Delta}|\chi^{(\alpha)}_{\mu_\alpha}\rangle, \quad
|\dot{\chi}_{\mu_\alpha}^{(\alpha)}\rangle=\hat{X}|\chi^{(\alpha)}_{\mu_\alpha}\rangle.
\eeq
Then, the variation and time derivative of total state are
compactly given by\cite{Miranda:2011a,Sato1,Sato2},
\begin{align}\label{defgf1}
|\delta\Psi\rangle = \sum_{\pmb{I}}\delta C_{\pmb{I}}|\pmb{I}\rangle +
\hat{\Delta}|\Psi\rangle, \quad |\dot{\Psi}\rangle =
\sum_{\pmb{I}}\dot{C}_{\pmb{I}}|\pmb{I}\rangle +
\hat{X}|\Psi\rangle, 
\end{align}
and their Hermitian conjugate are
\begin{align}\label{defgf2}
\langle\delta\Psi| =\sum_{\pmb{I}}\delta C^*_{\pmb{I}}\langle\pmb{I}| -
\langle\Psi|\hat{\Delta}, \quad \langle\dot{\Psi}|
=\sum_{\pmb{I}}\dot{C}^*_{\pmb{I}}\langle\pmb{I}| -
\langle\Psi|\hat{X}. 
\end{align}


It follows from Eq.(\ref{Var1}) that,
\begin{align}
\delta S &= \int_{t_0}^{t_1}\de t\left[\langle\delta\Psi|(H-\im\del_t)|\Psi\rangle
+ \langle\Psi|(H-\im\del_t)|\delta\Psi\rangle\right] \nonumber \\
&= \int_{t_0}^{t_1}\de t\langle\delta\Psi|\left[H|\Psi\rangle-\im|\dot{\Psi}\rangle\right] + \left[\langle\Psi|H + \im\langle\dot{\Psi}\right]|\delta\Psi\rangle.
\end{align}
Substituting Eqs.(\ref{defgf1}) and (\ref{defgf2}) into this equation,
after some algebraic manipulation \cite{Miranda:2011a,Sato1}, we obtain,
\begin{align}
\nonumber\label{Temp2}
\delta S &= \int\de t\sum_{\pmb{I}} \delta C_{\pmb{I}}^*\left\{\langle
\pmb{I}|H-\im\hat{X}|\Psi\rangle-\im\dot{C}_{\pmb{I}}\right\} \nonumber \\
&- \sum_{\pmb{I}}\left\{\langle\Psi|H-\im\hat{X}|\pmb{I}\rangle+\im\dot{C}^*_{\pmb{I}}\right\}\delta C_{\pmb{I}}\nonumber \\
&+\langle\Psi|\hat{\Delta}(1-\Pi)(\hat{H}-\im\hat{X})|\Psi\rangle \nonumber \\
&-\langle\Psi|(\hat{H}-\im\hat{X})(1-\Pi)\hat{\Delta}|\Psi\rangle,
\end{align}
where $\Pi = \sum_{\pmb{I}}|\pmb{I}\rangle\langle\pmb{I}|$ denotes the
projector onto the CI space, i.e., the subspace of $\pmb{N}$-electron Hilbert
space spanned by the configurations included in Eq.~(\ref{eq:MCTDHF ansatz}).
The action functional $S$ should be made stationary with respect to all
independent variations; $\{\delta{C}_{\pmb{I}},\delta{C}^*_{\pmb{I}}\}$
for CI coefficiens and $\{(\Delta_\alpha)^{\mu_\alpha}_{\nu_\alpha}\}$
for spin-orbitals. 
%

First, the EOM for CI coefficients are obtained from $\delta S/\delta C^*_{\pmb{I}}=0$, 
\beq\label{EOMforCI2}
\im\dot{C}_{\pmb{I}} =
\sum_{\pmb{J}}\langle\pmb{I}|\left(\hat{H}-i\hat{X}\right)|\pmb{J}\rangle
C_{\pmb{J}}.
\eeq
Requiring $\delta S/\delta C_{\pmb{I}}=0$ derives the complex conjugate
of Eq.~(\ref{EOMforCI2}). Next
from $\delta S/\delta (\Delta_\alpha)^{\mu_\alpha}_{\nu_\alpha}=0$, one
obtains
\begin{align}\nonumber\label{EOMforOrb1}
&\im\sum_{\beta}\sum_{\kappa_\beta\tau_\beta}\langle\Psi|\left[(\hat{E}_\alpha)^{\mu_\alpha}_{\nu_\alpha}
\bar{\Pi}(\hat{E}_\beta)^{\kappa_\beta}_{\tau_\beta}-(\hat{E}_\beta)^{\kappa_\beta}_{\tau_\beta}\bar{\Pi}(\hat{E}_\alpha)^{\mu_\alpha}_{\nu_\alpha}\right]|\Psi\rangle
(X_\beta)^{\kappa_\beta}_{\nu_\beta} 
\\ 
& \quad\quad = \langle\Psi|\left[(\hat{E}_\alpha)^{\mu_\alpha}_{\nu_\alpha}\bar{\Pi}\hat{H}-\hat{H}\bar{\Pi}(\hat{E}_\alpha)^{\mu_\alpha}_{\nu_\alpha}\right]|\Psi\rangle,
\end{align}
where $\bar{\Pi}=1-\Pi$. 
Equation~(\ref{EOMforOrb1}) 
is to be solved for $(X_\alpha)^{\mu_\alpha}_{\nu_\alpha} =
\langle\chi^{(\alpha)}_{\mu_\alpha}|\dot{\chi}^{(\alpha)}_{\nu_\alpha}\rangle$, thus
determines the time derivative of spin-orbitals.
We now take a closer look at Eq.~(\ref{EOMforOrb1}) for the following two distinct cases:\\

\noindent\underline{\textit{Case 1:}} $(\mu_\alpha,\nu_\alpha)=(i_\alpha,j_\alpha)$. In this
case we focus on the components of the spin-orbital variations
within the subspace spanned by the occupied spin-orbitals. Since $\bar{\Pi}(\hat{E}_\alpha)^{i_\alpha}_{j_\alpha}|\pmb{I}\rangle
\neq 0$ and $\langle\pmb{I}|(\hat{E}_\alpha)^{i_\alpha}_{j_\alpha}\bar{\Pi}\neq 0$ in general, 
one needs to directly work with
Eq.~(\ref{EOMforOrb1}) within the occupied spin-orbital space
\beq
&&\im\sum_{\beta}\sum_{k_\beta l_\beta}\langle\Psi|\left[(\hat{E}_\alpha)^{i_\alpha}_{j_\alpha}
\bar{\Pi}(\hat{E}_\beta)^{k_\beta}_{l_\beta}-(\hat{E}_\beta)^{k_\beta}_{l_\beta}\bar{\Pi}(\hat{E}_\alpha)^{i_\alpha}_{j_\alpha}\right]|\Psi\rangle
(X_\beta)^{k_\beta}_{j_\beta} 
\nonumber \\ 
&& \quad\quad = \langle\Psi|\left[(\hat{E}_\alpha)^{i_\alpha}_{j_\alpha}\bar{\Pi}\hat{H}-\hat{H}\bar{\Pi}(\hat{E}_\alpha)^{i_\alpha}_{j_\alpha}\right]|\Psi\rangle.
\label{eq:EOM-occupied}
\eeq
In the full-CI case, where
$\bar{\Pi}(\hat{E}_\alpha)^{i_\alpha}_{j_\alpha}|\Psi\rangle = 0$, $\langle\Psi|(\hat{E}_\alpha)^{i_\alpha}_{j_\alpha}\bar{\Pi} = 0$,
Eq.~(\ref{eq:EOM-occupied}) reduces to an identity $0=0$. Therefore, the corresponding $(X_\alpha)^{i_\alpha}_{j_\alpha}$ may be
arbitrary anti-Hermitian matrix elements, of which the simplest choice is $(X_\alpha)^{i_\alpha}_{j_\alpha} = 0$.\\

\noindent\underline{\textit{Case 2:}} $(\mu_\alpha,\nu_\alpha)=(i_\alpha,a_\alpha)$. In this
case we deal with the components of the spin-orbital
variations outside the occupied spin-orbital space. Since $\langle
\Psi|(\hat{E}_\alpha)^{i_\alpha}_{a_\alpha}\bar{\Pi} = \langle
\Psi|(\hat{E}_\alpha)^{i_\alpha}_{a_\alpha}$ and
$(\hat{E}_\alpha)^{i_\alpha}_{a_\alpha}|\Psi\rangle = 0$, Eq.~(\ref{EOMforOrb1}) becomes,
\begin{align}\label{temp4}
\im\sum_{\beta}\sum_{\kappa_\beta}\sum_{j_\beta}\langle\Psi|(\hat{E}_\alpha)^{i_\alpha}_{a_\alpha}(\hat{E}_\beta)^{\kappa_\beta}_{j_\beta}|\Psi\rangle
(X_\beta)^{\kappa_\beta}_{j_\beta} =
\langle\Psi|(\hat{E}_\alpha)^{i_\alpha}_{a_\alpha}\hat{H}|\Psi\rangle.
\end{align}
However, the matrix element in the left-hand side of the above equation
survives only when
$\beta=\alpha$ and $\kappa_\alpha=a_\alpha \in \Omega^{vir}_\alpha$, namely
$\langle\Psi|(\hat{E}_\alpha)^{i_\alpha}_{a_\alpha}(\hat{E}_\beta)^{\kappa_\beta}_{j_\beta}|\Psi\rangle
=\delta^{\alpha}_{\beta}\delta^{a_\alpha}_{\kappa_\alpha}\langle\Psi|(\hat{E}_\alpha)^{i_\alpha}_{j_\alpha}|\Psi\rangle
=
\delta^{\alpha}_{\beta}\delta^{a_\alpha}_{\kappa_\alpha}(\rho_\alpha)_{i_\alpha}^{j_\alpha}$.
Thus Eq.~(\ref{temp4}) is simplified to
\beq\label{eq:EOM-virtual}
\im\sum_{j_\alpha}(X_\alpha)^{a_\alpha}_{j_\alpha}(\rho_\alpha)_{i_\alpha}^{j_\alpha} =
\langle\Psi|(\hat{E}_\alpha)^{i_\alpha}_{a_\alpha}\hat{H}|\Psi\rangle.
\eeq
The $\pmb{m}$-body Hamiltonian contribution to the RHS of Eq.~(\ref{eq:EOM-virtual}) is
evaluated as follows;
\begin{align}
& 
\langle\Psi|(\hat{E}_\alpha)^{i_\alpha}_{a_\alpha}\hat{H}_{\pmb{m}}|\Psi\rangle =
\sum_{\pmb{\mu}\pmb{\nu}}\langle\Psi|(\hat{E}_\alpha)^{i_\alpha}_{a_\alpha}\hat{E}^{\pmb{\mu}}_{\pmb{\nu}}|\Psi\rangle
(H_{\pmb{m}})^{\pmb{\mu}}_{\pmb{\nu}} \nonumber \\
&=
\sum_{\pmb{\mu}\pmb{\nu}}
\langle\Psi|(\hat{E}_1)^{\pmb{\mu}_1}_{\pmb{\nu}_1}\cdot\cdot(\hat{E}_\alpha)^{i_\alpha}_{a_\alpha}(\hat{E}_\alpha)^{\pmb{\mu}_\alpha}_{\pmb{\nu}_\alpha}\cdot\cdot(\hat{E}_K)^{\pmb{\mu}_K}_{\pmb{\nu}_K}|\Psi\rangle
(H_{\pmb{m}})^{\pmb{\mu}_1\cdot\cdot\pmb{\mu}_\alpha\cdot\cdot\pmb{\mu}_K}_{\pmb{\nu}_1\cdot\cdot\pmb{\nu}_\alpha\cdot\cdot\pmb{\nu}_K} \nonumber \\ 
&= m_{\alpha} \sum_{j_\alpha}\sum_{\pmb{k}^{[\alpha]}\pmb{l}^{[\alpha]}}
\langle\Psi|(\hat{E}_1)^{\pmb{k}_1}_{\pmb{l}_1}\cdot\cdot(\hat{E}_\alpha)^{i_\alpha\pmb{k}_\alpha}_{j_\alpha\pmb{l}_\alpha}\cdot\cdot(\hat{E}_K)^{\pmb{k}_K}_{\pmb{l}_K}|\Psi\rangle 
(H_{\pmb{m}})^{\pmb{k}_1\cdot\cdot
a_\alpha\pmb{k}_\alpha\cdot\cdot\pmb{k}_K}_{\pmb{l}_1\cdot\cdot
j_\alpha\pmb{l}_\alpha\cdot\cdot\pmb{l}_K} \nonumber \\
&= m_\alpha\sum_{j_{\alpha}}\sum_{\pmb{k}^{[\alpha]}\pmb{l}^{[\alpha]}}
\langle\Psi|\hat{E}^{i_\alpha\pmb{k}^{[\alpha]}}_{j_\alpha\pmb{l}^{[\alpha]}}|\Psi\rangle 
(H_{\pmb{m}})^{a_\alpha\pmb{k}^{[\alpha]}}_{j_\alpha\pmb{l}^{[\alpha]}} \nonumber \\
&=m_\alpha\sum_{j_{\alpha}}\sum_{\pmb{k}^{[\alpha]}\pmb{l}^{[\alpha]}}
(H_{\pmb{m}})^{a_\alpha\pmb{k}^{[\alpha]}}_{j_\alpha\pmb{l}^{[\alpha]}}
(\rho_{\pmb{m}})_{i_\alpha\pmb{k}^{[\alpha]}}^{j_\alpha\pmb{l}^{[\alpha]}}.
\end{align}
In the second line of the above equation, we note that the
matrix element survives when one and only one of the $m_\alpha$ creation
operators in $(\hat{E}_\alpha)^{\pmb{\mu}_\alpha}_{\pmb{\nu}_\alpha}$ refers to $a_\alpha \in \Omega^{vir}_\alpha$, and all
the others to the occupied spin-orbitals. All such cases
[$\mu_{\alpha ,p}=a_\alpha, \mu_{\alpha ,q\neq p} \in \Omega^{occ}_\alpha; 1
\leq p \leq m_\alpha$] give the same contribution since the phase
$(\mp)^{p-1}$ [$+$ ($-$) sign for bosons (fermions), arising in
(anti-)commuting the creation operators] is canceled by shifting the corresponding annihilation
operator $\nu_{\alpha ,p}$, and the Hamiltonian is symmetric for
interchange of particles of the same kind. The third line is
thus obtained after renaming summation variables,
where 
$\pmb{k}^{[\alpha]}=(\pmb{k}_1,\cdots\pmb{k}_\alpha,\cdots\pmb{k}_K)$ is the array of $m-1$ indices with $\pmb{k}_{\alpha}
= (k_{\alpha ,2}\cdots ,k_{\alpha ,m_\alpha})$ and 
$\pmb{k}_{\beta}= (k_{\beta ,1}\cdots ,k_{\beta ,m_\beta})$ for
$\beta\neq\alpha$, with $\pmb{l}^{[\alpha]}$ defined similarly. 
The fourth line introduces the short-hand notation for the
array of $m$ indices,
$\mu_\alpha \pmb{k}^{[\alpha]} =
(\pmb{k}_1,\cdots\mu_\alpha\pmb{k}_\alpha,\cdots\pmb{k}_K)$
($\mu_\alpha \pmb{l}^{[\alpha]}$ is defined similarly), and the fifth
line uses the definition of the $\pmb{m}$-body RDM, Eq.~(\ref{eq:mRDM}).

Now the RHS of Eq.~(\ref{eq:EOM-virtual}) is given by the sum over $\pmb{m}$,
\beq
\langle\Psi|(\hat{E}_\alpha)^{i_\alpha}_{a_\alpha}\hat{H}|\Psi\rangle
&=& m_\alpha \sum_{\pmb{m}}
\sum_{j_{\alpha}}\sum_{\pmb{k}^{[\alpha]}\pmb{l}^{[\alpha]}}
(H_{\pmb{m}})^{a_\alpha\pmb{k}^{[\alpha]}}_{j_\alpha\pmb{l}^{[\alpha]}}
(\rho_{\pmb{m}})_{i_\alpha\pmb{k}^{[\alpha]}}^{j_\alpha\pmb{l}^{[\alpha]}}
\nonumber \\
&=&\langle\chi^{(\alpha)}_{a_\alpha}|\cdot\sum_{j_{\alpha}}
(\mathbf{H}_\alpha)_{i_\alpha}^{j_\alpha}
|\chi^{(\alpha)}_{j_\alpha}\rangle,
\eeq
where $(\mathbf{H}_\alpha)_{i_\alpha}^{j_\alpha}$ is the effective
one-particle operator,
\begin{align}\label{eq:heff}
(\mathbf{H}_\alpha)_{i_\alpha}^{j_\alpha} =
\sum_{\pmb{m}}\frac{m_{\alpha}}{\prod_{\beta=1}^{K}m_{\beta}!}\sum_{\pmb{k}^{[\alpha]}\pmb{l}^{[\alpha]}}
(W_{\pmb{m}})^{\pmb{k}^{[\alpha]}}_{\pmb{l}^{[\alpha]}}
(\rho_{\pmb{m}})_{i_\alpha\pmb{k}^{[\alpha]}}^{j_\alpha\pmb{l}^{[\alpha]}},
\end{align}
and $(W_{\pmb{m}})^{\pmb{k}^{[\alpha]}}_{\pmb{l}^{[\alpha]}}$ is given in the
coordinate representation as
\begin{align}
&(W_{\pmb{m}})^{\pmb{k}^{[\alpha]}}_{\pmb{l}^{[\alpha]}}(x_\alpha,x^\prime_\alpha)
= \nonumber \\ 
&\int\de\pmb{y}^{[\alpha]}\de\pmb{z}^{[\alpha]}
\varphi^*_{\pmb{k}^{[\alpha]}}(\pmb{y}^{[\alpha]})H_{\pmb{m}}(x_\alpha \pmb{y}^{[\alpha]}, x^\prime_\alpha\pmb{z}^{[\alpha]})
\varphi_{\pmb{l}^{[\alpha]}}(\pmb{z}^{[\alpha]}),
\end{align}
where
$\pmb{y}^{[\alpha]}=(\pmb{y}_1,\cdots\pmb{y}_\alpha,\cdots\pmb{y}_K)$ is the set of $m-1$ coordinates with $\pmb{y}_\alpha
= (y_{\alpha ,2}\cdots ,y_{\alpha ,m_\alpha})$ and 
$\pmb{y}_{\beta}= (y_{\beta ,1}\cdots ,y_{\beta ,m_\beta})$ for
$\beta\neq\alpha$, and $x_{\alpha}\pmb{y}^{[\alpha]} =
(\pmb{y}_1,\cdots x_\alpha\pmb{y}_\alpha ,\cdots\pmb{y}_K)$ is the array
of $m$ coordinates. $\pmb{z}^{[\alpha]}$ and
$x^\prime_\alpha\pmb{z}^{[\alpha]}$ are defined similarly.

Finally, gathering the occupied
and virtual
components of the time
derivative completes the derivation of EOM for spin-orbitals
\beq\label{EOMforOrb}
&&\im|\dot{\chi}^{(\alpha)}_{i_\alpha}\rangle = \im\sum_{j_\alpha}|\chi^{(\alpha)}_{j_\alpha}\rangle
(X_\alpha)^{j_\alpha}_{i_\alpha} \nonumber\\ &&+
\sum_{a_\alpha}|\chi^{(\alpha)}_{a_\alpha}\rangle\langle\chi^{(\alpha)}_{a_\alpha}|\sum_{j_\alpha k_\alpha}
(\mathbf{H}_\alpha)^{j_\alpha}_{k_\alpha}|\chi^{(\alpha)}_{j_\alpha}\rangle(\rho^{-1}_\alpha)^{k_\alpha}_{i_\alpha} \\
&&= \im\sum_{j_\alpha}|\chi^{(\alpha)}_{j_\alpha}\rangle
(X_\alpha)^{j_\alpha}_{i_\alpha} +
(1-\hat{P}_\alpha)\sum_{j_\alpha k_\alpha}
(\mathbf{H}_\alpha)^{j_\alpha}_{k_\alpha}|\chi^{(\alpha)}_{j_\alpha}\rangle(\rho^{-1}_\alpha)^{k_\alpha}_{i_\alpha},\nonumber 
\eeq
where $\hat{P}_{\alpha} = \sum_{i_\alpha}|\chi_{i_\alpha}\rangle\langle\chi_{i_\alpha}|$
is the {spin-orbital} projection operator onto the occupied spin-orbital
space, with which the virtual space $\sum_{a_\alpha}|\chi^{(\alpha)}_{a_\alpha}\rangle\langle\chi^{(\alpha)}_{a_\alpha}|=1-\hat{P}_\alpha$ is referenced as
a whole, thus avoiding explicit use of virtual spin-orbitals.
$(X_\alpha)^{j_\alpha}_{i_\alpha}$ in the first term is to be obtained by solving Eq.~(\ref{eq:EOM-occupied}), and, as discussed above, can be set zero in the full-CI case.
Equation (\ref{EOMforCI2}) for CI coefficients and Eq.~(\ref{EOMforOrb}) for spin-orbitals form fully general TD-MCSCF equations of motion, not restricted to full CI, for a system composed of any arbitrary kinds and numbers of fermions and bosons. 

\section{Molecules interacting with an external laser field}
\label{sec:molecule}

In this Section we present the working equations for a molecule subject to an external laser field.
Let the molecule consist of electrons and $K_{\rm n}$ different kinds of
nuclei treated quantum mechanically (the kind does {\it not} necessarily corresponds to the
nuclear species, see discussion below), and $N_{\rm cl}$ nuclei treated as a classical point charge. For clarity and notational simplicity, we assign the electrons to the first
kind of particle ($\alpha = 1$), and kinds $\alpha = 2,3,\cdots ,K$ represent quantum nuclei with $K = 1+K_{\rm n}$.
The numbers of identical particles are, as before, denoted by
$\{N_\alpha\}$. Then the number of electrons is $N_1$, the number
of quantum nuclei is $N_{\rm n} = \sum_{\alpha =2}^K$, and the total
number of atoms is $N_{\rm atom} = N_{\rm n} + N_{\rm cl}$.
We use atomic units in this section.

The spin-independent molecular Hamiltonian in the coordinate representation is given by
\beq\label{eq:hmol}
&&H = \sum_{\alpha =1}^K\sum_{p_\alpha =1}^{N_\alpha} h_{\alpha}(\pmb{r}_{p_\alpha},\pmb{r}^\prime_{p_\alpha},t)
+ \sum_{\alpha =1}^K\sum_{p_\alpha =1}^{N_\alpha}\sum_{q_\alpha > p_\alpha}^{N_\alpha} U_{\alpha\alpha}(|\pmb{r}_{p_\alpha}-\pmb{r}_{q_\alpha}|) \nonumber \\
&&\quad+ \sum_{\alpha =1}^K\sum_{\beta > \alpha}^K\sum_{p_\alpha =1}^{N_\alpha}\sum_{q_\beta =1}^{N_\beta} U_{\alpha\beta}(|\pmb{r}_{p_\alpha}-\pmb{r}_{q_\beta}|),
\eeq
where
$U_{\alpha\beta}(r) = Z_\alpha Z_\beta/r$ is the Coulomb interaction
with $Z_\alpha$ being the electric charge, and 
\begin{align}\label{eq:hmol1}
h_{\alpha}(\pmb{r},\pmb{r}^\prime,t) = \delta(\pmb{r}-\pmb{r}^\prime)\left[
-\frac{\nabla^2_{\pmb{r}^\prime}}{2m_\alpha} 
+\sum_{A=1}^{N_{\rm cl}}\frac{Z_\alpha Z_A}{|\pmb{r} - \pmb{R}_A|}
\right]+ V^{\rm ext}_\alpha(\pmb{r},\pmb{r}^\prime,t),
\end{align}
is the one-particle Hamiltonian composed of the kinetic energy [the
first term with $m_\alpha$ being the mass (not to be confused with the
number of particles)], Coulomb interaction with classical nuclei with the
charges $\{Z_A\}$ located at $\{\pmb{R}_A\}$ (the second term), and the time-dependent laser-particle interaction
$V^{\rm ext}_\alpha$, 
given, e.g., within the dipole approximation either in
the length gauge (LG) or in the velocity gauge (VG), by
\begin{align}
V^{\rm ext}_{\alpha,{\rm LG}}(\pmb{r},\pmb{r}^\prime,t) &=
-\delta(\pmb{r}-\pmb{r}^\prime) Z_\alpha \pmb{E}(t)\cdot\pmb{r}, \\
V^{\rm ext}_{\alpha,{\rm VG}}(\pmb{r},\pmb{r}^\prime,t) &=
\delta(\pmb{r}-\pmb{r}^\prime) \im\frac{Z_\alpha}{m_\alpha}\pmb{A}(t)\cdot\pmb{\nabla}_{\pmb{r}^\prime}
+\frac{Z_\alpha^2}{2m_\alpha}|\pmb{A}(t)|^2,
\end{align}
where $\pmb{E}(t)$ is the laser electric field, and $\pmb{A}(t) = -\int
 \pmb{E}(t) \de t$ is the vector potential.

The general formulation of Sec.~\ref{sec:EOM} is readily applicable to the
molecular Hamiltonian of Eq.~(\ref{eq:hmol}). The CI EOM reads
\beq\label{eq:eommolci}
&&\im\dot{C}_{\pmb{I}} = \sum_{\pmb{J}}\langle
\pmb{I}|\left[\sum_\alpha\sum_{i_\alpha
j_\alpha}(\tilde{h}_\alpha)^{i_\alpha}_{j_\alpha}(\hat{E}_\alpha)^{i_\alpha}_{j_\alpha}
\right. \nonumber \\ &&\left.
\quad\quad\quad+\frac{1}{2}\sum_{\alpha}\sum_{i_\alpha j_\alpha k_\alpha l_\alpha}
(U_{\alpha\alpha})^{i_\alpha k_\alpha}_{j_\alpha l_\alpha} (\hat{E}_{\alpha})^{i_\alpha k_\alpha}_{j_\alpha l_\alpha}
\right. \\ &&\left.
\quad\quad\quad+\frac{1}{2}\sum_{\alpha}\sum_{\beta\ne\alpha}\sum_{i_\alpha j_\alpha k_\beta l_\beta}
(U_{\alpha\beta})^{i_\alpha k_\beta}_{j_\alpha l_\beta} 
(\hat{E}_{\alpha})^{i_\alpha}_{j_\alpha}
(\hat{E}_{\beta})^{k_\beta}_{l_\beta}
\right]|\pmb{J}\rangle C_{\pmb{J}},\nonumber 
\eeq
where
\begin{align}\label{eq:h1matmol}
(\tilde{h}_\alpha)^{i_\alpha}_{j_\alpha} = \int \de x_\alpha
\chi^{(\alpha)*}_{i_\alpha}(x_\alpha) [h_{\alpha}(t) \chi^{(\alpha)}_{j_\alpha}](x_\alpha) - i({X_\alpha})^{i_\alpha}_{j_\alpha},
\end{align}
\begin{align}
&(U_{\alpha\beta})^{i_\alpha k_\beta}_{j_\alpha l_\beta} \nonumber\\
&= Z_\alpha Z_\beta
\int\de x_\alpha\de x^\prime_\beta
\frac{\chi^{(\alpha)*}_{i_\alpha}(x_\alpha)\chi^{(\beta)*}_{k_\beta}(x^\prime_\beta)\chi^{(\alpha)}_{j_\alpha}(x_\alpha)\chi^{(\beta)}_{l_\beta}(x^\prime_\beta)}
{|\pmb{r}_\alpha-\pmb{r}^\prime_\beta|},
\label{eq:h2matmol}
\end{align}
with $x_\alpha = (\pmb{r}_\alpha ,\sigma_\alpha)$ being the composite spatial- and spin-coordinates, 
and the EOM for spin-orbitals is given by
\begin{align}\label{eq:eommolorb}
\im|\dot{\chi}^{(\alpha)}_{i_\alpha}\rangle =
\im\sum_{j_\alpha}|\chi^{(\alpha)}_{j_\alpha}\rangle ({X_\alpha})^{j_\alpha}_{i_\alpha}
+ (1-\hat{P}_\alpha) \left(h_{\alpha} + V^{(\alpha)}_{i_\alpha}\right)|\chi^{(\alpha)}_{i_\alpha}\rangle,
\end{align}
where the one-body contribution 
to the second term of
Eq.~(\ref{EOMforOrb}) is extracted to lead to $h_\alpha|\chi^{(\alpha)}_{i_\alpha}\rangle$ by noting 
$[(\rho_\alpha)(\rho_\alpha)^{-1}]^{i_\alpha}_{j_\alpha}=\delta^{i_\alpha}_{j_\alpha}$, and
\begin{align}\label{eq:eommolorb}
V^{(\alpha)}_{i_\alpha}|\chi^{(\alpha)}_{i_\alpha}\rangle =
\sum_{j_\alpha j^\prime_\alpha}
\sum_\beta\sum_{k_\beta
l_\beta}(W_{\alpha\beta})^{k_\beta}_{l_\beta}|\chi^{(\alpha)}_{j^\prime_\alpha}\rangle
(\rho_{\alpha\beta})_{j_\alpha k_\beta}^{j^\prime_\alpha l_\beta}
(\rho^{-1}_\alpha)_{i_\alpha}^{j_\alpha},
\end{align}
\beq
(W_{\alpha\beta})^{k_\beta}_{l_\beta}(x_\alpha) =
Z_\alpha Z_\beta \int\de x_\beta^\prime \frac{\chi^{(\beta)*}_{k_\beta}(x_\beta^\prime)\chi^{(\beta)}_{l_\beta}(x_\beta^\prime)}
{|\pmb{r}_\alpha - \pmb{r}_\beta^\prime|}.
\eeq

Finally, Eq.~(\ref{eq:EOM-occupied}) is formulated as the linear system
of equations,
\beq\label{eq:eommolorbocc}
\sum_{\beta}\sum_{k_\beta l_\beta} (A_{\alpha\beta})^{i_\alpha k_\beta}_{j_\alpha
l_\beta} (X_\beta)^{k_\beta}_{l_\beta} = (B_\alpha)^{i_\alpha}_{j_\alpha}.
\eeq
where $(A_{\alpha\beta})^{i_\alpha k_\beta}_{j_\alpha l_\beta} =
(\bar{A}_{\alpha\beta})^{i_\alpha k_\beta}_{j_\alpha l_\beta} - (\bar{A}_{\beta\alpha})^{k_\beta i_\alpha}_{l_\beta j_\alpha}$,
$(B_\alpha)^{i_\alpha}_{j_\alpha}=(F_\alpha)^{i_\alpha}_{j_\alpha}-(F_\alpha)^{j_\alpha *}_{i_\alpha}$, with
\begin{align}\label{eq:amat}
(\bar{A}_{\alpha\beta})^{i_\alpha k_\beta}_{j_\alpha l_\beta} = \delta^{\alpha}_{\beta}
(\rho_\alpha)^{l_\alpha}_{i_\beta}\delta^{j_\alpha}_{k_\alpha} -
\langle\Psi|(\hat{E}_\alpha)^{i_\alpha}_{j_\alpha}\Pi(\hat{E}_\alpha)^{k_\beta}_{l_\beta}|\Psi\rangle,
\end{align}
\begin{align}\label{eq:fvec}
&(F_\alpha)^{i_\alpha}_{j_\alpha} = \sum_{k_\alpha}
(h_\alpha)^{j_\alpha}_{k_\alpha}(\rho_\alpha)_{i_\alpha}^{k_\alpha}  + \sum_{k_\alpha}
\sum_\beta\sum_{k_\beta l_\beta} 
(U_{\alpha\beta})^{j_\alpha k_\beta}_{k_\alpha l_\beta}
(\rho_{\alpha\beta})_{i_\alpha k_\beta}^{k_\alpha l_\beta}
\nonumber \\ 
&\quad\quad-\langle\Psi|(\hat{E}_\alpha)^{i_\alpha}_{j_\alpha}\Pi\hat{H}|\Psi\rangle.
\end{align}
In order for Eq.~(\ref{eq:eommolorbocc}) to be solvable
(with non-singular coefficient matrix $A$), one needs a systematic method of constructing non-full-CI space analogous to the TD-ORMAS method \cite{Sato2} for electrons.
We shall discuss this issue in the future publication.

Equations of motions~(\ref{eq:eommolci}) and (\ref{eq:eommolorb}), with
the matrix equation (\ref{eq:eommolorbocc}) 
defines the general TD-MCSCF method, not restricted to full CI, for molecules interacting with an
external field. Our formulation is very flexible; it includes as
special cases both the electron dynamics at the classical-nuclei approximation ($N_{\n}=0,
N_{\rm cl}=N_{\rm atom}$) and the full quantum molecular dynamics
($N_{\n}=N_{\rm atom},N_{\rm cl}=0$). 
Furthermore, it allows various approaches to
the same physical problem; e.g., the same nuclear species in the
molecule can be treated either as identical particles or distinguishable ones to
investigate the physical outcomes of the particle statistics during the
course of laser-molecule interaction.\\

\section{Summary}
\label{sec:summary}

We have developed a fully general {\it ab initio} TD-MCSCF approach to describe the dynamics of a many-body system that is a mixture of any arbitrary kinds and numbers of fermions and bosons subject to an external field. In this approach, the total wave function is expanded in terms of configurations constructed from time-dependent single-particle spin-orbitals. The expansion is not limited to the full-CI one, and the configurations used in the expansion can be specified in terms of the whole mixture rather than each particle kind separately. The equations of motion for the CI coefficients and spin-orbitals have been derived, based on the time-dependent variational principle. Furthermore, we have presented the working equations applicable to investigation of the ultrafast dynamics in a molecule irradiated by intense laser fields and/or ultrashort XUV pulses.

The present framework is highly flexible. For example, we can treat identical nuclei in spatially separated subdomains of a molecule as different particle kinds. We can also treat heavy nuclei and incident projectiles as classical particles instead of quantum ones. The latter may be handled as an external field as well.

Whereas our original motivation lies in {\it ab initio} simulations of the electron-nuclear dynamics in molecules driven by a laser pulse, our method will be applicable to a wide variety of problems far beyond. Especially, the Hamiltonian can contain non-local terms and involve many-body (more than two-body) interactions. Thus, it may also find applications in cold-atom/cold-molecule physics and nuclear physics.


\begin{acknowledgments}

We thank Joachim Burgd\"orfer for helpful discussions.
This research was supported in part by a Grant-in-Aid for Scientific Research (Grants No.~25286064, No.~26390076, No.~26600111, and No.~16H03881) from the Ministry of Education, Culture, Sports, Science and Technology (MEXT) of Japan and also by the Photon Frontier Network Program of MEXT.
This research was also partially supported by the Center of Innovation Program from the Japan Science and Technology Agency, JST, and by CREST (Grant No.~JPMJCR15N1), JST.
R.A. gratefully acknowledges support from the Graduate School of Engineering, The University of Tokyo, Doctoral Student Special Incentives Program (SEUT Fellowship).

\end{acknowledgments}





\bibliography{rsc}

\end{document}